\documentclass[aip,jcp, reprint,groupedaddress]{revtex4-2}
\usepackage{graphicx}
\usepackage{tabularx}
\usepackage{dcolumn}
\usepackage{longtable}
\usepackage{tensor}
\usepackage{color}
\usepackage{placeins}
\usepackage{bm}
\usepackage{amsmath}
\usepackage{amsfonts}
\usepackage{amssymb}
\usepackage{float}
\usepackage[urlcolor=white]{hyperref}
\providecommand{\U}[1]{\protect\rule{.1in}{.1in}}
\graphicspath{{./Figures/}}

\begin{document}

\title{Nonzero-temperature vibronic spectra of polyatomic molecules from a zero-temperature classical trajectory}
\author{Davide Barbiero}
\author{Ji\v{r}\'i J. L. Van\'i\v{c}ek}
\email{jiri.vanicek@epfl.ch}
\affiliation{Laboratory of Theoretical Physical Chemistry, Institut des Sciences et
Ing\'enierie Chimiques, Ecole Polytechnique F\'ed\'erale de Lausanne (EPFL),
CH-1015, Lausanne, Switzerland}
\date{\today}

\begin{abstract}
By combining coherence thermofield dynamics with the single-Hessian
approximation, we enable simulations of low- to medium-resolution vibronic
spectra of weakly anharmonic systems at nonzero temperatures, at negligible
additional cost relative to zero-temperature calculations. Single-Hessian
coherence thermofield Gaussian wavepacket dynamics is exact in any harmonic
potential, provided that the reference Hessian is that of the final surface.
When applied to Morse systems of increasing anharmonicity and varying
temperature, this method successfully captures excited-state anharmonicity
and key temperature-dependent spectral features, including hot bands and
broadening. By combining the method with on-the-fly ab initio dynamics, we
demonstrate its utility by computing the absorption spectra of naphthalene,
aminocoumarin C450, and phenyl radical, and the photoelectron spectrum of $%
\text{SeO}_{2}^{-}$. Within the ab initio single-Hessian approximation,
after the zero-temperature spectrum is obtained at the cost of classical
molecular dynamics (on the order of hours), all nonzero-temperature spectra
are computed in seconds.
\end{abstract}

\maketitle


\section{Introduction \label{sec:introduction}}

Vibrationally resolved electronic spectroscopy provides insights into
biochemical processes, atmospheric chemistry, and material design.~\cite%
{Barone_Puzzarini:2021,book_Marquardt_Quack:2021_v2} Since experiments are
often performed at room or higher temperatures,~\cite%
{Jaiswal_Nenov:2024,Zhang_Gogotsi:2023,Zou_Li:2020,Timmer_Lienau:2022,Zhou_Peng:2023} including thermal
effects in theoretical models is crucial for reliable interpretation and
prediction of spectra.~\cite%
{Xun_He:2020,Wernbacher_Gonzalez:2021_v2,Prlj_Curchod:2022}
State-of-the-art approaches employ either time-independent sum-over-states
expressions~\cite{Huh_Aspuru-Guzik:2015,Barone:2016,Ru_Sanov:2022} or
Fourier transform of time correlation functions obtained from (i)~direct
evolution of the thermal density matrix,~\cite%
{Reddy_Prasad:2016_v2,Ren_Chan:2018,Gherib_Ryabinkin:2022} (ii)~statistical
sampling,~\cite{Richardson_Thoss:2017,Karsten_Kuhn:2018} or (iii)~ensemble
averaging correlation functions of pure quantum states.~\cite%
{Srsen_Heger:2020,Dral_Barbatti:2021}

In coherence thermofield dynamics,~\cite{Begusic_Vanicek:2020} the
evaluation of the coherence component of the density matrix is transformed
exactly to the calculation of a wavepacket autocorrelation function in an
augmented configuration space. Although initially derived in harmonic
systems,~\cite{Ritschel_Eisfeld:2015,Reddy_Prasad:2015} coherence
thermofield dynamics is fully general;~\cite{Begusic_Vanicek:2020} this
powerful simplification enables simulation of spectra at nonzero
temperatures using any method for zero-temperature wavepacket dynamics.
However, exact quantum propagation of the wavepacket~\cite%
{Zhang_Vanicek:2024} is feasible only for small systems due to exponential
scaling with the doubled number of degrees of freedom and the requirement of
a global potential energy surface (PES). Consequently, harmonic models
remain widely used, both because they are readily available in commercial
software and because errors of the electronic structure often outweigh
inaccuracies in nuclear dynamics. Going beyond harmonic models~\cite{Aieta_Ceotto:2020,Moscato_Ceotto:2024} does not
require highly accurate wavefunctions: spectra are typically recorded at low
resolution and are therefore dominated by short-time nuclear dynamics.~\cite%
{Heller:1981a} In this regime, simple methods such as the single-trajectory
thawed Gaussian approximation can already yield reliable spectra, despite
neglecting quantum effects like tunneling or wavepacket splitting.~\cite%
{Heller:1975} Combination of the coherence thermofield approach with the
thawed Gaussian and extended thawed Gaussian approximations has enabled
accurate calculations of both linear~\cite%
{Begusic_Vanicek:2020,Begusic_Vanicek:2021a} and nonlinear~\cite%
{Begusic_Vanicek:2021} vibronic spectra as well as quantum yields~\cite%
{Wenzel_Mitric:2023a} in mildly anharmonic systems. Notably,
nonzero-temperature effects are included with nearly no additional
computational cost relative to the zero-temperature thawed Gaussian
approximation.~\cite{Begusic_Vanicek:2020}

However, the thawed Gaussian approximation---based on a local harmonic
approximation of the potential and henceforth referred to as local harmonic
Gaussian wavepacket dynamics (GWD)---still requires evaluation of the Hessian
of the molecular potential along the trajectory,~\cite%
{Scheidegger_Golubev:2022,Kletnieks_Vanicek:2023,Gherib_Genin:2024_v2}
hindering on-the-fly applications. To speed up the local harmonic GWD in the
zero-temperature setting, Begu\v{s}i\'{c}, Cordova, and Van\'i\v{c}ek
proposed the single-Hessian GWD,~\cite{Begusic_Vanicek:2019,Vanicek:2023_v2}
which relies on the evaluation of only one Hessian for the entire trajectory
and, despite its simplicity, captures anharmonicity in zero-temperature
vibronic spectra with accuracy comparable to that of the local harmonic
approach.~\cite%
{Begusic_Vanicek:2019,Prlj_Vanicek:2020,Begusic_Vanicek:2022_v2,Barbiero_Vanicek:2026}%

Here, we combine single-Hessian GWD and coherence thermofield dynamics to
obtain a cost-effective tool for on-the-fly ab initio prediction of
vibronic spectra at arbitrary temperatures. After deriving the equations
needed for numerical propagation of the wavepacket, we demonstrate how
spectra at nonzero temperatures can be obtained from a single
zero-temperature classical trajectory. Having validated the approach in
Morse model systems (by comparison with exact quantum calculations), we
demonstrate its practical utility in on-the-fly ab initio simulations of the
photoelectron spectrum of $\text{SeO}_{2}^{-}$ and of the absorption spectra
of phenyl radical, naphthalene, and aminocoumarin C450.


\section{Results and discussion\label{sec:results}}

\subsection{Single-Hessian coherence thermofield Gaussian wavepacket
dynamics \label{subsec:theory}}

Within the electric dipole approximation, first-order time-dependent
perturbation theory, and Condon approximation, the rotationally averaged
absorption cross-section from the ground state $|g\rangle$ to an excited
state $|e\rangle$ can be computed as the Fourier transform 
\begin{equation}
\sigma(\omega)=\frac{4\pi\omega}{3\hbar c}\mu^{2}\text{Re}%
\int_{0}^{\infty}C(t)\,e^{i\omega t}\,dt   \label{eq:spectra}
\end{equation}
of the electronic coherence~\cite{book_Tannor:2007} 
\begin{equation}
C(t)=\text{Tr}\left(e^{-i\hat{H}_{e}t/\hbar}\hat{\rho}\,e^{i\hat{H}%
_{g}t/\hbar}\right)  \label{eq:coherence}
\end{equation}
between states $|g\rangle$ and $|e\rangle$. Here, $\mu=\lVert\vec{\mu}%
_{eg}\rVert$ is the magnitude of the transition dipole moment, $\hat{\rho}$
is the temperature-dependent vibrational density operator in the ground
electronic state, and 
\begin{equation}
\hat{H}_{j}=V_{j}(\hat{q})+T(\hat{p}), \,\, j=g\,\text{or}\,e,
\end{equation}
are the vibrational Hamiltonians in the two electronic states. These
Hamiltonians depend on an associated adiabatic PES $V_{j}(\hat{q})$ and on
the kinetic energy operator $T(\hat{p})=\hat{p}^{T} \cdot m^{-1} \cdot \hat{p%
}/2$, where $m$ is a $D \times D$ mass matrix and $\hat{p}$ is the $D$%
-component momentum operator.

In coherence thermofield dynamics, the coherence~(\ref{eq:coherence}) is
rewritten exactly as the wavepacket autocorrelation function~\cite%
{Begusic_Vanicek:2020} 
\begin{equation}
C(t)=\langle\bar{\psi}_{0}|\bar{\psi}_{t}\rangle  \label{eq:autocorr}
\end{equation}
of the initial thermofield state 
\begin{equation}
\bar{\psi}_{0}(\bar{q})=\langle q|\hat{\rho}^{1/2}|q^{\prime}\rangle
\label{eq:psi_CTF}
\end{equation}
propagated according to the time-dependent Schr\"{o}dinger equation 
\begin{equation}
i \hbar \dot{\bar{\psi}}_{t}=\hat{\bar{H}}\bar{\psi}_{t},  \label{eq:TDSE}
\end{equation}
with augmented Hamiltonian $\bar{H}(\bar{q})=H_{e}(q)-H_{g}(q^{\prime})$ in
the $2D$-dimensional augmented configuration space $\bar{q}=(q,q^{\prime})$,
consisting of the ``physical'' ($q$) and ``fictitious'' ($q^{\prime}$)
degrees of freedom. The thermofield wavepacket thus encodes coherence
between dynamics on the two PESs.

Following Ref.~\onlinecite{Begusic_Vanicek:2020} (but using the ``$QPS$''
parameterization~\cite{Heller:1976a_v2,Hagedorn:1980,Lasser_Lubich:2020}), we
approximate this wavepacket with a Gaussian 
\begin{align}
\bar{\psi}_{t}(\bar{q})&= (\pi\hbar)^{-D/2} (\text{det} \, \bar{Q}%
_{t})^{-1/2}\nonumber\\
&~~~\times\text{exp}\bigg[\frac{i}{\hbar}\bigg(\frac{1}{2}\,\bar{x}^{T}
\cdot \bar{P}_{t} \cdot \bar{Q}^{-1}_{t}\cdot \bar{x} +\bar{p}_{t}^{T} \cdot 
\bar{x}+\bar{S}_{t}\bigg)\bigg],  \label{eq:GWP}
\end{align}
where $\bar{x}:=\bar{q}-\bar{q}_{t}$ is the shifted position vector, $\bar{q}%
_{t}$ and $\bar{p}_{t}$ are the position and momentum of the Gaussian's
center, $\bar{Q}_{t}$ and $\bar{P}_{t}$ are two complex-valued $2D$%
-dimensional matrices that determine the symmetric width matrix $\bar{A}%
_{t}:=\bar{P}_{t}\cdot \bar{Q}_{t}^{-1}$ of the Gaussian, and $\bar{S}_{t}$
is a real scalar. At time $t=0$, the Gaussian wavepacket is centered at the
equilibrium geometry of the ground-state surface in both the physical and
fictitious coordinates ($q_{0}=q^{\prime}_{0}=q_{g,\text{eq}}$), with zero
initial momentum ($p_{0}=p_{0}^{\prime}=0$). The width of the Gaussian is
determined by the vibrational thermal density matrix associated with the
harmonic fit at the minimum of the PES of the electronic ground state. In
particular, $\bar{Q}_{0}=\bar{\Gamma}^{-1/2}$ and $\bar{P}_{0}=i\bar{\Gamma}%
^{1/2}$, with 
\begin{align}
\bar{\Gamma}&=%
\begin{pmatrix}
C & S \\ 
S & C%
\end{pmatrix}
,  \label{eq:gamma}
\end{align}
$C=m^{1/2}\cdot\Omega\cdot\text{coth}(\Omega_{\beta})\cdot m^{1/2}$, $%
S=-m^{1/2}\cdot\Omega\cdot\text{csch}(\Omega_{\beta})\cdot m^{1/2}$, $%
\Omega=(m^{-1/2}\cdot \kappa_{g} \cdot m^{-1/2})^{1/2}$, $%
\Omega_{\beta}=\beta\hbar\Omega/2$, and $\kappa_{g}:=\text{Hess}\,V_{g}(q_{g,%
\text{eq}})$.

To reduce the computational burden of the local harmonic approach from Ref.~%
\onlinecite{Begusic_Vanicek:2020}, we adopt the single-Hessian approximation 
\begin{align}
\bar{V}_{\text{SHA}}(\bar{q};\bar{q}_{t}) & := \bar{V}(\bar{q}_{t})+\text{%
grad}\,\bar{V}(\bar{q}_{t})^{T} \cdot \bar{x} + \bar{x}^{T} \cdot \bar{\kappa%
}_{\text{ref}} \cdot \bar{x}/2  \label{eq:V_SHA}
\end{align}
for the augmented potential $\bar{V}(\bar{q})=V_{e}(q)-V_{g}(q^{\prime})$.
Inserting Eqs.~(\ref{eq:GWP}) and (\ref{eq:V_SHA}) into Eq.~(\ref{eq:TDSE})
yields 
\begin{align}
\dot{\bar{q}}_{t} & =\bar{m}^{-1}\cdot \bar{p}_{t},  \label{eq:qEOM_SHA} \\
\dot{\bar{p}}_{t} & =-\text{grad}\,\bar{V}_{\text{SHA}}(\bar{q}_{t})=-\text{%
grad}\,\bar{V}(\bar{q}_{t}),  \label{eq:pEOM_SHA} \\
\dot{\bar{Q}}_{t} & =\bar{m}^{-1}\cdot \bar{P}_{t},  \label{eq:QEOM_SHA} \\
\dot{\bar{P}}_{t} & =-\text{Hess}\,\bar{V}_{\text{SHA}}(\bar{q}_{t})\cdot 
\bar{Q}_{t}=-\bar{\kappa}_{\text{ref}}\cdot \bar{Q}_{t},  \label{eq:PEOM_SHA}
\\
\dot{\bar{S}}_{t} & =\bar{T}(\bar{p}_{t})-\bar{V}_{\text{SHA}}(\bar{q}_{t})=%
\bar{T}(\bar{p}_{t})-\bar{V}(\bar{q}_{t}),   \label{eq:SEOM_SHA}
\end{align}
with $\bar{T}(\bar{p}_{t})=T(p_{t})-T(p^{\prime}_{t})$. The augmented mass
and reference ``single'' Hessian matrices are the direct sums $\bar{m}%
=m\oplus(-m)$ and $\bar{\kappa}_{\text{ref}}=\kappa_{\text{ref}%
}\oplus(-\kappa_{g})$. 
Three natural choices $\kappa_{\text{ref}}=\text{Hess}\,V_{e}(q_{e,\text{eq}%
})$, $\text{Hess}\,V_{e}(q_{g,\text{eq}})$, and $\text{Hess}\,V_{g}(q_{g,%
\text{eq}})$ correspond to so-called adiabatic, vertical, and initial
single-Hessian approximations and are discussed in Refs.~\onlinecite%
{Begusic_Vanicek:2019,Barbiero_Vanicek:2026} for the zero-temperature case.

The dynamics of the center $(\bar{q}_{t},\bar{p}_{t})$ of the Gaussian is
independent of temperature. Equations~(\ref{eq:qEOM_SHA}) and~(\ref%
{eq:pEOM_SHA}) are solved by propagating two independent trajectories in $D$
dimensions; trajectory $(q_{t},p_{t})$ is propagated with $H_{e}$, while
trajectory $(q_{t}^{\prime},p_{t}^{\prime})$ is propagated with the negative
of $H_{g}$ (or equivalently, backward in time with $H_{g}$). Since the
second trajectory starts at the minimum of the ground-state PES with zero
momentum, it remains stationary. Thus, only a single trajectory on the
excited-state surface is required. Remarkably, this single trajectory $%
(q_{t},p_{t})$ is the same at all temperatures and does not have to be
recomputed after the zero-temperature spectrum has been evaluated.

In contrast, the dynamics of the width $\bar{A}_{t}:=\bar{P}_{t}\cdot \bar{Q}%
_{t}^{-1}$ of the Gaussian [Eqs.~(\ref{eq:QEOM_SHA}) and~(\ref{eq:PEOM_SHA}%
)] does depend on temperature. Since the matrices $\bar{m}$ and $\bar{\kappa}%
_{\text{ref}}$ are block-diagonal, the $2D\times 2D$ problem of propagating $%
(\bar{Q}_{t},\bar{P}_{t})$ reduces to four independent $D\times D$ blocks
and can be solved exactly to give quasi-periodic oscillations of the width.
Using a constant Hessian for propagating the width may sound crude but has
several advantages: the single-Hessian method (i)~still includes
anharmonicity because it uses the exact anharmonic classical trajectory $(%
\bar{q}_{t},\bar{p}_{t})$, (ii)~circumvents the nonphysical, unbounded
growth of the width amplitude observed in the local harmonic approach,~\cite%
{Ryabinkin_Genin:2024} and (iii)~conserves the effective energy and
symplectic structure, and nearly conserves the energy of the wavepacket.~\cite%
{Barbiero_Vanicek:2026} Most importantly, the spectrum at any temperature
can be computed at the cost of one ab initio classical trajectory. Finally,
single-Hessian GWD remains exact for many-dimensional displaced, distorted,
and Duschinsky-rotated harmonic potentials if $\kappa_{\text{ref}}$ is based
on the Hessian at any position of the excited-state surface.


\subsection{Morse potential \label{subsec:Morse}}

To test the accuracy of the proposed method, we constructed a
one-dimensional model consisting of a ground-state harmonic potential $%
V_{g}(q)=q^{T}\cdot\kappa_{g}\cdot q/2$ and an anharmonic excited-state
Morse potential 
\begin{equation}
V_{e}(q)=V_{e,\text{eq}}+\frac{\omega}{4\chi}[1-e^{-\sqrt{2m\omega\chi}%
(q-q_{e,\text{eq}})}]^{2}.  \label{eq:Morse}
\end{equation}
The parameters ($\kappa_{g}$, $V_{e,\text{eq}}$, $\omega$, $\chi$, $m$, and $%
q_{e,\text{eq}}$) were chosen to reproduce the $\tilde{\text{A}}\leftarrow%
\tilde{\text{X}}$ transition in BaS.~\cite{Huber_Herzberg:1979_v2}
Assuming a harmonic ground-state surface ensures that any discrepancy from
the exact quantum results arises solely from the approximate treatment of
the anharmonic excited-state dynamics. 
Spectra were evaluated at scaled temperatures $T_{\omega}=1/(\beta\hbar%
\Omega)\in\{0,0.5,1\}$.

\begin{figure*}
\centering
\includegraphics{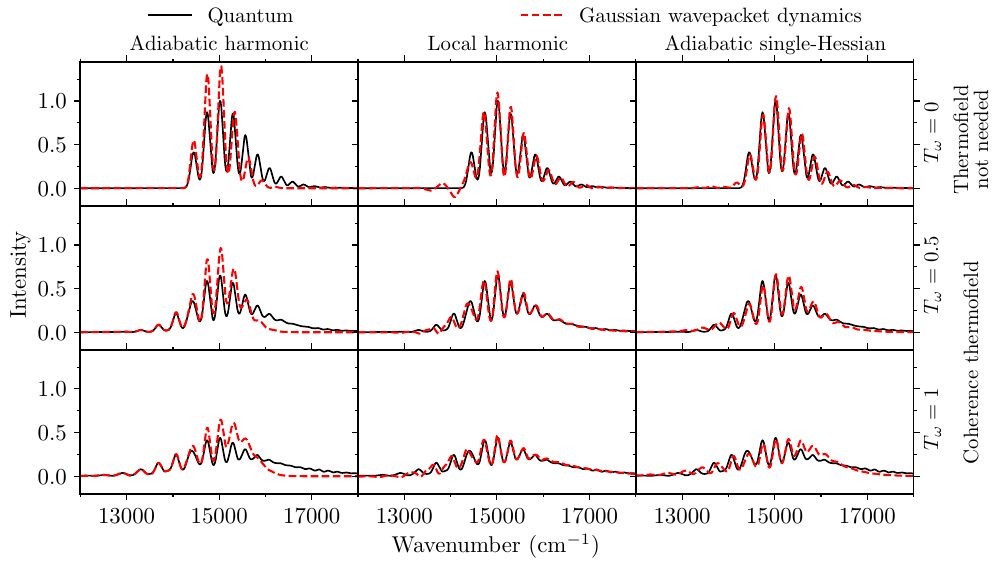} 
\caption{Vibronic spectra of BaS at three different scaled temperatures $T_{%
\protect\omega}$ obtained from the split-operator coherence thermofield
dynamics and different effective potentials for the semiclassical coherence
thermofield Gaussian wavepacket dynamics. The anharmonicity of the
excited-state surface~(\protect\ref{eq:Morse}) of BaS is described by
the dimensionless anharmonicity parameter $\protect\chi%
\approx0.01$ and the Huang-Rhys factor $S\approx2.4$.}
\label{fig:Morse_spec}
\end{figure*}

Figure~\ref{fig:Morse_spec} compares spectra computed from adiabatic
single-Hessian coherence thermofield Gaussian wavepacket dynamics with those
obtained from exact quantum,~\cite{Zhang_Vanicek:2024} adiabatic harmonic,
and local harmonic~\cite{Begusic_Vanicek:2020} coherence thermofield
dynamics. At zero temperature, the adiabatic harmonic model describes
reasonably well the frequency and relative (but not absolute) intensity of
transitions to low-excited vibrational states near the minimum of the PES
but loses accuracy for transitions to higher-excited, more anharmonic
vibrational states. Compared with the global harmonic approximation, local
harmonic and single-Hessian GWD substantially improve the description of the
frequency, and of both the relative and absolute intensity of the peaks.
However, small, nonphysical negative peaks appear in the spectra, due to the
nonlinear character of the single-Hessian and (especially) local-harmonic
effective Hamiltonians.~\cite{Vanicek:2023_v2,Barbiero_Vanicek:2026}

As the temperature increases, hot bands emerge and the spectral range
widens. The adiabatic harmonic model accurately captures hot bands below the
0-0 transition, corresponding to transitions to final states with low
vibrational quantum number. However, the accuracy of this model deteriorates
compared to the zero-temperature case in the high-frequency region of the
spectrum because the increasingly delocalized thermal wavepacket is unable
to explore the strongly anharmonic regions of the final-state PES. In
contrast, local harmonic GWD accurately describes thermally activated
transitions to highly excited, anharmonic vibrational states but is less
reliable at frequencies below the 0-0 transition, where hot bands overlap
with the nonphysical negative peaks. Single-Hessian GWD provides a middle
ground: it achieves accuracy comparable to that of local harmonic
GWD while avoiding both the spurious negative peaks at low wavenumbers and
the high computational cost of evaluating Hessians along the trajectory.

In Fig. S1 of the supporting information, we further probe the performance of the methods when exploring anharmonic regions of the PES. The single-Hessian coherence thermofield GWD outperforms the harmonic approximation for all anharmonicities and at all temperatures studied. Although highly reliable at low temperatures, the single-Hessian approximation seems to deteriorate slightly faster than the local harmonic one as the temperature increases.


\subsection{$\mathbf{\text{SeO}_{2} +e^{-} \leftarrow \text{SeO}_{2}^{-}}$
photoelectron spectrum at 700 K\label{subsec:SeO2}}

Coherence thermofield dynamics is particularly well suited for anion
photoelectron spectroscopy, where elevated temperatures are often required
to generate sufficient ion densities.~\cite%
{Snodgrass_Bowen:1989,Ru_Sanov:2022} Expressions from Sec.~\ref%
{subsec:theory} apply unchanged to anion photoelectron spectroscopy, if $%
|g\rangle$ and $|e\rangle$ are reinterpreted as the ionic and neutral ground
states, respectively.

The experimental photoelectron spectrum of $\text{SeO}_{2}^{-}$ features two
hot bands,~\cite{Snodgrass_Bowen:1989} marked by asterisks in Fig.~\ref%
{fig:SeO2_hot_bands}. Single-Hessian coherence thermofield GWD captures not
only these experimentally observed hot bands, but also additional thermally
activated transitions that overlap with fundamentals and enhance the
spectral baseline. In the experiment, the ion source setup
precluded the acquisition of reliable data above $18000\,\text{cm}^{-1}$.~%
\cite{Snodgrass_Bowen:1989}

\begin{figure}
\includegraphics{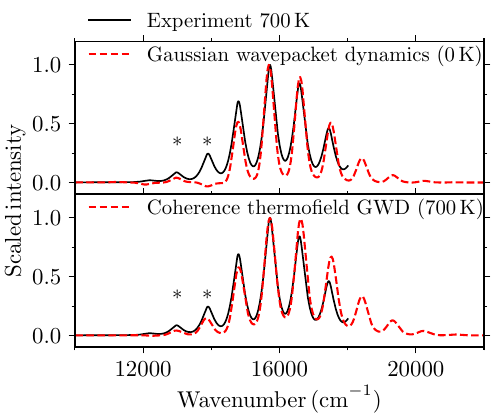} %
\caption{Photoelectron spectrum of $\text{SeO}_{2}^{-}$ at $700\,\text{K}$.
Spectra computed from the vertical single-Hessian Gaussian wavepacket
dynamics (GWD) at $0\,\text{K}$ and from the vertical single-Hessian
coherence thermofield GWD at $700\,\text{K}$ are compared to the
experimental spectrum.~\cite{Snodgrass_Bowen:1989} Hot bands are denoted by
asterisks.} \label{fig:SeO2_hot_bands}
\end{figure}

The single-Hessian approach again substantially outperforms the
global harmonic model constructed from the same Hessian and achieves
accuracy comparable to that of the computationally much more demanding local
harmonic approach in simulating the photoelectron spectrum of $\text{SeO}%
_{2}^{-}$ (see Fig.~\ref{fig:SeO2_anharmonicity}). The global harmonic model
does not reproduce the correct peak spacing because it cannot capture the
anharmonicity of the final-state PES. In contrast, both local harmonic and
single-Hessian methods reproduce the spacing accurately because the
corresponding wavepackets evolve along the same anharmonic classical guiding
trajectory. Despite their similar accuracy, the computational cost of the
two approaches differs markedly. At zero temperature, single-Hessian GWD
required $4.8\times 10^{3}\,\text{s}\approx 1.3\,\text{h}$ compared to $%
2.2\times 10^{4}\,\text{s}\approx 6.1\,\text{h}$ for local harmonic GWD
using analytical Hessians; employing numerical Hessians would increase the
cost of local harmonic GWD to $8.2\times 10^{4}\,\text{s}\approx 22.8\,\text{%
h}$. All calculations were performed on 8 cores. Once the ab initio
trajectory needed for the zero-temperature spectrum is computed, the
spectrum at $700\,\text{K}$ is obtained in under one second.

\begin{figure}
\includegraphics{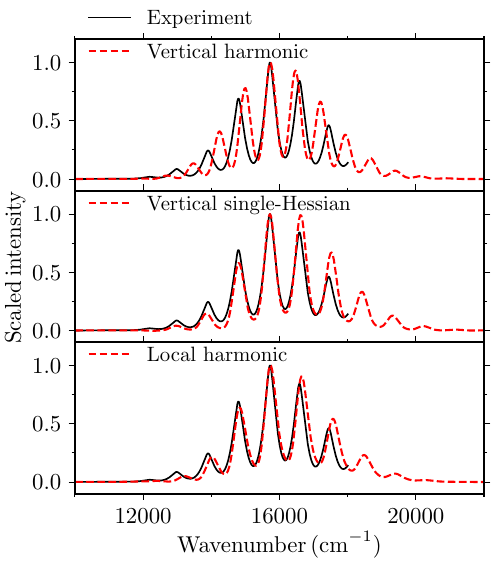} %
\caption{Photoelectron spectrum of $\text{SeO}_{2}^{-}$ at $700\,\text{K}$.
Spectra computed from the vertical harmonic, vertical single-Hessian, and
local harmonic coherence thermofield Gaussian wavepacket dynamics are
compared to the experimental spectrum.~\cite{Snodgrass_Bowen:1989}} \label%
{fig:SeO2_anharmonicity}
\end{figure}

The relatively long vibrational periods of $\text{SeO}_{2}$ (see the
harmonic wavenumbers in Table S2 of the supporting information) allowed
us to propagate the Gaussian wavepacket~(\ref{eq:GWP}) using a large time
step and therefore to exploit the advantages of high-order geometric
integrators.~\cite{Barbiero_Vanicek:2026} In particular, we employed a
fourth-order integrator, which requires no extra coding beyond the
second-order integrator,~\cite{Barbiero_Vanicek:2026} yet achieves faster
convergence of the spectrum, as demonstrated in Sec.~S1 of the supporting
information.


\subsection{Room-temperature absorption spectrum of phenyl radical\label%
{subsec:Phenyl}}

\begin{figure}
\includegraphics{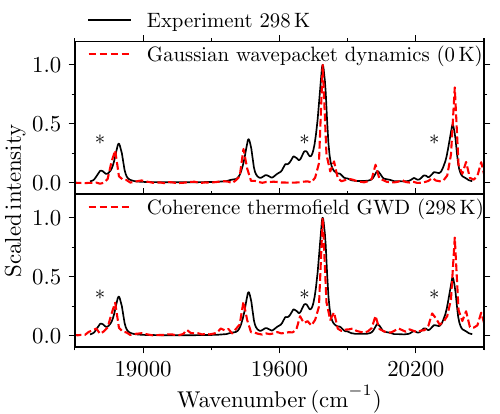} \caption{Absorption
spectrum of phenyl radical at $298\,\text{K}$. Spectra computed from the
adiabatic single-Hessian Gaussian wavepacket dynamics (GWD) at $0\,\text{K}$
and from the adiabatic single-Hessian coherence thermofield GWD at
$298\,\text{K}$ are compared to the experimental
spectrum.~\cite{Tonokura_Wallington:2002} Hot bands are denoted by
asterisks.} \label{fig:Phe_spec}
\end{figure}

Figure~\ref{fig:Phe_spec} compares the low-energy region of the absorption
spectrum of phenyl radical computed at zero temperature from the GWD and at
room temperature from the coherence thermofield GWD (both based on the
adiabatic single-Hessian approximation) to the experimental spectrum
measured at room temperature.~\cite{Tonokura_Wallington:2002} To facilitate
comparison with the rotationally averaged computed spectra, the rotational
structure was removed from the experimental spectrum by broadening (see
Fig.~S3 of the supporting information). The inclusion of temperature
effects via coherence thermofield dynamics allows single-Hessian GWD to
capture the three hot bands (marked by asterisks in Fig.~\ref{fig:Phe_spec})
qualitatively. The predicted spectral region up to $30000\,\text{cm}^{-1}$
is presented in Fig.~S5 of the supporting information.

\subsection{Temperature-dependent absorption spectrum of naphthalene\label%
{subsec:Naph}}

In Secs.~\ref{subsec:Morse}-\ref{subsec:Phenyl}, we presented spectra in
which the vibrational structure was sufficiently resolved to reveal hot
bands. In practice, however, electronic spectra are often recorded at lower
resolution, and therefore the dominant effect of temperature becomes an
overall broadening of the spectrum.

\begin{figure*}
\includegraphics{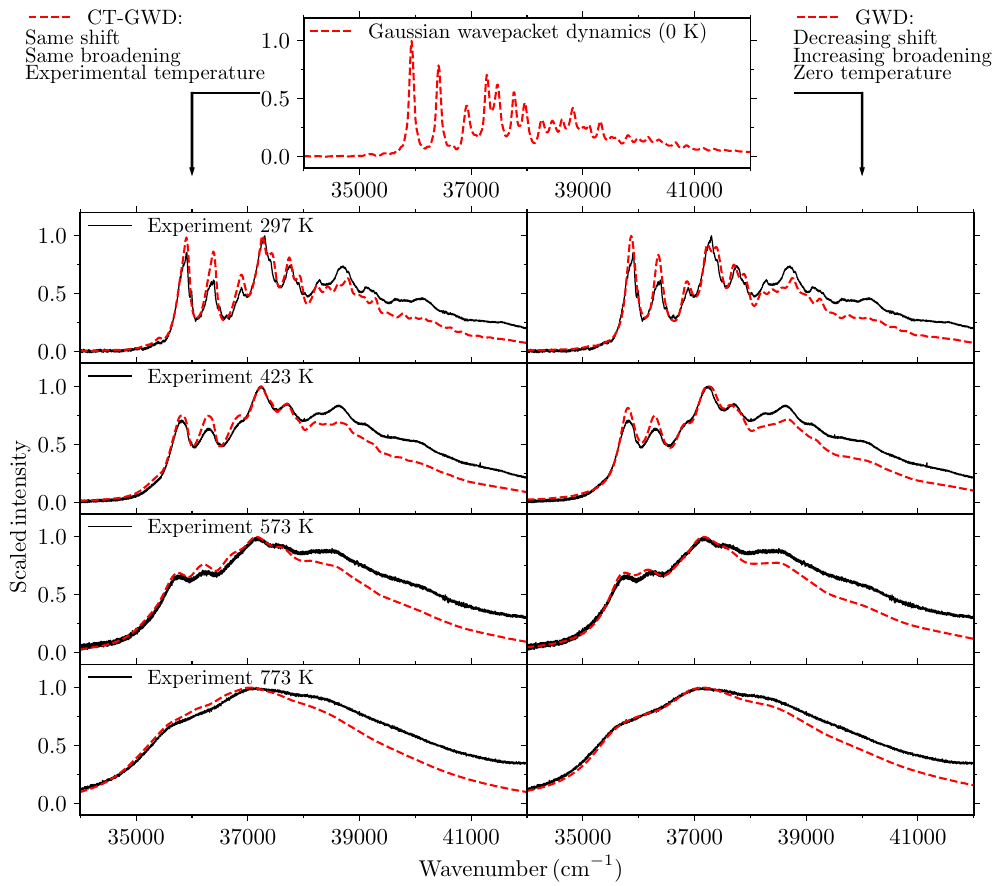} 
\caption{Absorption spectra of naphthalene at different temperatures. The spectrum computed from the adiabatic single-Hessian Gaussian wavepacket dynamics (GWD) at $0\,\text{K}$ is shown in the top middle panel. 
The left column compares experimental spectra~\cite{Grosch_Fateev:2015} with spectra computed from adiabatic single-Hessian coherence thermofield (CT) GWD at the corresponding temperatures, using the same spectral broadening and shift for all temperatures. 
The right column compares the experimental spectra with spectra computed from adiabatic single-Hessian GWD at $0\,\text{K}$, where the broadening and shift are fitted ad hoc for each temperature.
} \label{fig:naph_spec}
\end{figure*}

Here, we focus on naphthalene, for which the absorption cross-section at
four different temperatures is available.~\cite{Grosch_Fateev:2015} In such
cases, it is commonly assumed that an ad hoc, artificial broadening of the
spectrum computed at zero temperature is sufficient to reproduce the
experimental spectrum accurately.~\cite%
{Benkyi_Sundholm:2019,Chadwick_Besley:2020,Kose:2021} Although
this strategy performs reasonably well (see Fig.~\ref{fig:naph_spec}), it is
used mainly because conventional approaches for incorporating temperature
effects are computationally expensive, especially when each temperature
requires a separate simulation. In addition, the broadened
zero-temperature spectra must be shifted independently for each temperature,
which is another ad hoc adjustment without any physical basis because the
shift should only be needed to correct a systematic error due to the
electronic structure method, which obviously does not depend on temperature.
Within single-Hessian coherence thermofield GWD, the spectra at the four
different temperatures---as well as at any other temperature of interest---can
be computed directly from the ab initio data needed already for the
zero-temperature spectrum, without requiring additional electronic structure
evaluations. For naphthalene, classical evolution of the nuclei at zero
temperature required $3.8\times 10^{5}\,\text{s}\approx 4.4\,\text{d}$ on 8
cores, whereas propagation of the width required less than $10\,\text{s}$
for each temperature of interest. While remaining as inexpensive
as broadening the zero-temperature spectrum, single-Hessian coherence
thermofield GWD avoids the need for separate ad hoc adjustments for
each temperature. Figure~\ref{fig:naph_spec} shows that the
coherence thermofield approach naturally captures the temperature-dependent
broadening of the spectrum. Moreover, spectra at different temperatures are
accurate despite using a common shift for all temperatures.

Increasing the temperature enhances the contribution of transitions to
higher excited electronic states in the high-frequency region of the
spectrum. Neglect of these contributions in our simulations explains the
slightly reduced accuracy of computed spectra at higher temperatures (see
also the absolute cross-sections in Fig.~S7 of the supporting
information).

\subsection{Room-temperature absorption spectrum of aminocoumarin C450\label%
{subsec:C450}}

Because broadening the zero-temperature single-Hessian spectrum
is an ad hoc procedure for including nonzero-temperature effects, it can
lead to incorrect merging of neighboring spectral peaks. Although equally
inexpensive, single-Hessian coherence thermofield GWD incorporates
temperature effects in a more physical way and can therefore remain reliable
even in such challenging cases.

\begin{figure}
\includegraphics{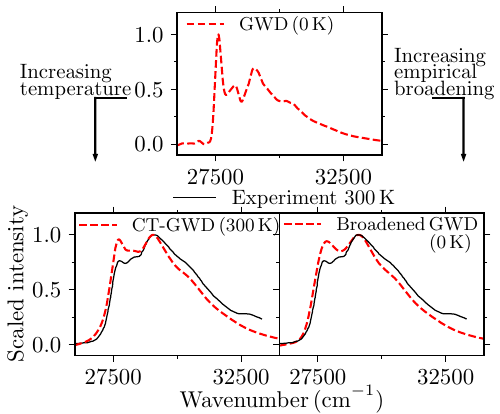} \caption{Absorption
spectrum of aminocoumarin C450 at $300\,\text{K}$. For comparison with the
experimental spectrum,~\cite{MunizMiranda_Barone:2015} the spectrum computed
from the adiabatic single-Hessian Gaussian wavepacket dynamics (GWD) at
$0\,\text{K}$ (top panel) can be post-processed either by including
temperature effects via coherence thermofield (CT) dynamics (bottom left
panel) or by increasing the empirical broadening (bottom right panel).} %
\label{fig:C450_spec}
\end{figure}

For example, the experimental spectrum~\cite{MunizMiranda_Barone:2015} of 
aminocoumarin C450 in methylcyclohexane features three peaks
between 27500 and 30000 $\text{cm}^{-1}$, along with a shoulder at higher
frequency (see Fig.~\ref{fig:C450_spec}). However, broadening the
zero-temperature spectrum results in the merging of the first two peaks. In
contrast, the spectrum computed from coherence thermofield dynamics captures
the three-peak progression.

Finally, Fig.~S8 of the supporting information highlights the importance of
accounting for anharmonicity even in simulation of low-resolution spectra.
For different reasons,~\cite{Begusic_Vanicek:2019,Barbiero_Vanicek:2026}
both vertical and adiabatic harmonic calculations fail to reproduce the
shape of the experimental spectrum. In contrast, such shape is captured by
the single-Hessian method, independently of the choice (adiabatic or
vertical) of the reference Hessian.


\section{Conclusion \label{sec:conclusion}}

We introduced single-Hessian coherence thermofield Gaussian wavepacket
dynamics for the efficient computation of vibronic spectra at nonzero
temperatures. This method combines single-Hessian Gaussian wavepacket
dynamics~\cite{Begusic_Vanicek:2019} with the thermofield-wavepacket
representation of electronic coherence.~\cite%
{Begusic_Vanicek:2020,Zhang_Vanicek:2024} The coherence thermofield
representation bypasses the need for Boltzmann averaging, statistical
sampling, and solving the von Neumann equation. Instead, the method solves
the time-dependent Schr\"{o}dinger equation for a Gaussian wavepacket in a
phase space of doubled dimension. Due to the single-Hessian approximation,
the evolution of the Gaussian only requires the evolution of a single
classical trajectory on the final-state surface, along with the harmonic
(i.e., analytically solvable) evolution of the width. Because the width
dynamics is inexpensive and the trajectory is temperature-independent,
spectra at any temperature can be computed from a single ab initio classical
molecular dynamics simulation, i.e., from the classical evolution of the
nuclei in the phase space.

In our numerical examples, we first studied one-dimensional Morse potential
systems with varying extent of anharmonicity and at different temperatures.
Spectra obtained from the single-Hessian coherence thermofield GWD always
outperformed those from global harmonic models; moreover, the single-Hessian
approximation showed accuracy comparable to that of the more expensive local
harmonic approximation. These results establish the single-Hessian coherence
thermofield GWD as a cost-effective and robust approach for simulating low-
to medium-resolution spectra of weakly anharmonic molecules, particularly
when the molecules are large or require high-level description of electronic
structure.

Then, we investigated the vibronic spectra of four different molecular
systems at several temperatures. Although single-Hessian coherence
thermofield GWD offers a general framework for simulating various linear and
nonlinear electronic spectroscopies, we tested the method specifically
against experimental data from linear absorption and photoelectron
spectroscopies. Emission and fluorescence spectra were not treated because
these techniques are often compromised by inefficient vibrational relaxation
following optical excitation, leading to hot-band features that deviate from
thermal equilibrium and obscure direct comparison with theory. In the
systems studied here ($\text{SeO}_{2}$, phenyl radical, naphthalene, and
aminocoumarin C450), a very good agreement with experimental spectra over a
wide range of temperatures confirms that the single-Hessian coherence
thermofield GWD successfully captures moderate excited-state anharmonicity
(see also Figs.~S2,~S4, and~S6 of the supporting information) and key
temperature-dependent spectral features, including hot bands and spectral
broadening.

This method is highly efficient, requiring only hours to compute the
zero-temperature spectrum and seconds to obtain spectra at any additional
temperature. Although its single-trajectory nature prevents the
applicability of single-Hessian coherent-state thermofield GWD in cases
where quantum effects such as tunneling or wavepacket splitting play a
dominant role in shaping the spectrum, this method offers a computationally
affordable improvement over harmonic models, which remain the standard
approach for predicting vibronic spectra. Together, these features suggest
that the method is broadly applicable across a wide range of weakly
anharmonic systems.

To facilitate the application of the method, we
provide a user-friendly Python implementation. The program requires only a
zero-temperature classical trajectory---readily obtainable from standard ab
initio molecular dynamics packages---and a reference Hessian matrix as input.
From these minimal ingredients, the program constructs the Gaussian
wavepacket, performs the analytical evolution of the width, and computes
both the wavepacket autocorrelation function and the resulting vibronic
spectrum at any desired temperature. By lowering the technical barrier to
implementation, this tool makes the method both accessible and highly
practical for routine applications.


\section{Methods \label{sec:details}}

\subsection{Morse potential}

The wavepacket was propagated for $10000$ steps with a time step $\Delta
t=8\, \text{a.u.}$ (i.e., for a total time of $80000 \, \text{a.u.}$) in the
adiabatic single-Hessian, adiabatic harmonic, and local harmonic effective
potentials, using a fourth-order geometric integrator,~\cite%
{book_Lubich:2008,Vanicek:2023_v2,Barbiero_Vanicek:2026} or with quantum
dynamics using the analogous fourth-order split operator algorithm.~\cite%
{Roulet_Vanicek:2019,Zhang_Vanicek:2024} The position grid for quantum
dynamics consisted of $256$ equidistant points between $-100$ and $900$ in
the physical dimension and of $32$ equidistant points between $-125$ and $125
$ in the fictitious dimension. A Gaussian damping function with a half-width
at half-maximum of $4000\, \text{a.u.}$ was applied to each autocorrelation
function before computing the spectrum.


\subsection{On-the-fly ab initio calculations}

Simulations were performed using our in-house Fortran 2018 code \texttt{%
molequle} for molecular quantum dynamics, interfaced with Gaussian 16
package~\cite{Frisch_Fox:2016} for on-the-fly evaluations of electronic
structure. Ground-state geometries were optimized using the density
functional theory, while the time-dependent density functional theory was
used for calculations in excited electronic states. The functionals were
B3LYP (Ref.~\onlinecite{Becke:1993}) for naphthalene, $\omega$B97X-D (Ref.~\onlinecite%
{Chai_Head-Gordon:2008}) for aminocoumarin C450,~\cite%
{MunizMiranda_Barone:2015} BPW91 (Ref.~\onlinecite{Perdew_Wang:1996}) for $\text{%
SeO}_{2}$,~\cite{Fu_Liang:2022} and CAM-B3LYP (Ref.~\onlinecite{Yanai_Handy:2004})
for phenyl. The augmented correlation-consistent polarized valence
quadruple-zeta (aug-cc-pVQZ) basis set~\cite{Kendall_Harrison:1992_v2} was
used for $\text{SeO}_{2}$, and the 6-31+G** basis set~\cite%
{Frisch_Binkley:1984} for all other systems. For C450, solvent effects were
not taken into account.~\cite{MunizMiranda_Barone:2015} See the
supporting information for the optimized geometries of all species and
for the harmonic vibrational wavenumbers of $\text{SeO}_{2}$ and phenyl
radical.

Geometric integrators based on a splitting approach were used to solve the
differential equations~(\ref{eq:qEOM_SHA})-(\ref{eq:SEOM_SHA}) of
single-Hessian GWD.~\cite{Barbiero_Vanicek:2026} We employed the
fourth-order integrator with a time step of $80 \, \text{a.u.}$ for $\text{%
SeO}_{2}$, and the second-order integrator with a time step of $8 \, \text{%
a.u.}$ for the other systems. The temperature-independent propagation of
trajectories on the excited-state surface was performed in normal-mode
coordinates, using the equilibrium geometry on the PES of the electronic
ground state as reference Eckart frame. For phenyl, the wavepacket was
propagated for $80000 \, \text{a.u.} \approx 1953.10\, \text{fs}$. For
naphthalene, the propagation time was $16000 \, \text{a.u.} \approx 387.02\, 
\text{fs}$ and the correlation functions were padded with zeros up to $%
80000\, \text{a.u.}$ to obtain smooth spectra in the frequency domain. For
selenium dioxide and C450, the propagation time was $8000 \, \text{a.u.}
\approx 193.51\, \text{fs}$ and the correlation functions were padded with
zeros up to $40000\, \text{a.u.}$ When included for comparison, local
harmonic GWD was performed using integrators equivalent to those employed in
the single-Hessian calculations,~\cite{Vanicek:2023_v2} while global
harmonic GWD was always performed using the exact propagation scheme
described in Ref.~\onlinecite{Barbiero_Vanicek:2026}.

The computed spectra were broadened using Lorentzian functions with a half
width at half maximum of $8.45\, \text{cm}^{-1}$ for phenyl and $50\, \text{%
cm}^{-1}$ for naphthalene, and Gaussian functions with a half width at half
maximum of $100\, \text{cm}^{-1}$ for aminocoumarin C450 and $170\, \text{cm}%
^{-1}$ for $\text{SeO}_{2}$. To account for the error in the accuracy of the
electronic structure, wavenumber scaling factors were employed. These
factors were 0.952 for aminocoumarin C450 and 0.963 for naphthalene and
phenyl radical.~\cite{cccbdb_2019} Computed spectra were shifted to match
the most intense experimental peak; the shifts are listed in Tables~S3, S7,
S10, and~S14 of the supporting information.


\section*{Data availability}

The Python program and the data that support the findings of this study are
openly available in Zenodo at http://doi.org/10.5281/zenodo.17779461.

\section*{Supporting Information}

See the supporting information for (i)~analysis of the accuracy of single-Hessian coherence thermofield GWD with respect to two dimensionless parameters characterizing anharmonicity of a Morse potential, (ii)~optimized geometries, harmonic
spectra, and spectral shifts of naphthalene, aminocoumarin C450, $\text{SeO}%
_{2}$, and phenyl radical, (iii)~comparison of the rotationally resolved and
broadened experimental spectra of phenyl radical, (iv)~prediction of the
vibronic spectrum of phenyl radical in the $20500-30000\,\text{cm}^{-1}$
range, and (v)~the computed absolute cross-sections for naphthalene.

\section*{Acknowledgments} 

The authors thank K. Tonokura for sharing the raw experimental data of
phenyl absorption and \={E}. Kl\={e}tnieks for performing preliminary
calculations for aminocoumarin C450. This research was supported by the
Swiss National Science Foundation (Grant No. 10005187) and by EPFL.

\bibliographystyle{aipnum4-2}
\bibliography{Single_Hessian_thermofield_v34}

\end{document}


\title{Supporting information for ``Nonzero-temperature vibronic spectra of polyatomic molecules from a zero-temperature classical trajectory''}

\author{Davide Barbiero}
\author{Ji\v{r}\'i J. L. Van\'i\v{c}ek}
\email{jiri.vanicek@epfl.ch}

\affiliation{Laboratory of Theoretical Physical Chemistry, Institut des Sciences et Ing\'enierie Chimiques, Ecole Polytechnique F\'ed\'erale de Lausanne (EPFL), CH-1015, Lausanne, Switzerland}

\date{\today}

\begin{abstract}
This document, which provides supporting information to the main text, contains (i)~analysis of the accuracy of single-Hessian coherence thermofield GWD with respect to two dimensionless parameters characterizing anharmonicity of a Morse potential, (ii)~optimized geometries, harmonic spectra, and spectral shifts of naphthalene, aminocoumarin C450, $\text{SeO}_{2}$, and phenyl radical, (iii)~comparison between the rotationally resolved and broadened experimental spectra of phenyl radical, (iv)~prediction of the vibronic spectrum of phenyl radical in the 20500-30000 $\text{cm}^{-1}$ range, and (v)~the computed absolute cross-sections of naphthalene.
\end{abstract}

\maketitle


\section{Morse potential}

In Fig. 1 of the main text, we observed that single-Hessian coherence thermofield Gaussian wavepacket dynamics (GWD) outperforms the global harmonic model and achieves accuracy similar to that of the local harmonic version in computing the absorption spectrum of BaS.
To further probe the performance of the methods when exploring anharmonic
regions of the PES, we systematically varied anharmonicity by modifying
either the dimensionless anharmonicity parameter $\chi $ or the Huang-Rhys
factor $S=m\Omega q_{e,\text{eq}}^{2}/(2\hbar )$, dimensionless
parameter characterizing the displacement between the two surfaces. To
compare the three approximate methods quantitatively, in Fig.~\ref%
{fig:Morse_anhar} we measured the error of an approximate spectrum with the
spectral contrast angle $\theta $ between the approximate and exact spectra.
This angle is defined through its cosine as~\cite{Begusic_Vanicek:2019} 
\begin{equation}
\text{cos}\,\theta =\frac{\sigma _{\text{ref}}\cdot \sigma }{\lVert \sigma _{%
\text{ref}}\rVert \,\lVert \sigma \rVert },  \label{eq:spec_angle}
\end{equation}%
where $\sigma _{1}\cdot \sigma _{2}:=\int \sigma _{1}(\omega )\sigma
_{2}(\omega )d\omega $ is the inner product of two spectra and $\lVert
\sigma \rVert =(\sigma \cdot \sigma )^{1/2}$ is the associated norm. The
single-Hessian coherence thermofield GWD outperforms the harmonic
approximation for all anharmonicities and at all temperatures studied.
Although highly reliable at low temperatures, the single-Hessian
approximation seems to deteriorate slightly faster than the local harmonic
one as the temperature increases.

\begin{figure}
\includegraphics{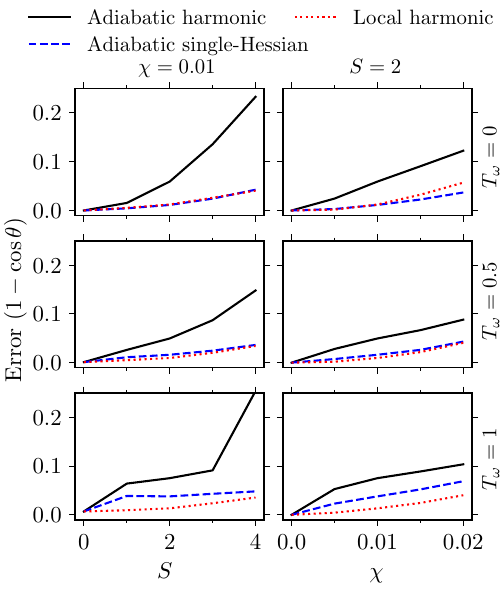} 
\caption{Errors, measured by the spectral contrast angle $\theta$ [see
Eq.~(\ref{eq:spec_angle}) for the definition], of the spectra computed with
the different semiclassical approaches in a Morse
potential~[Eq.~(15) of the main text], as a function of the Huang-Rhys factor $S$ for a
fixed anharmonicity parameter $\chi=0.01$ (left panels), and as a function
of the anharmonicity parameter $\chi$ for a fixed Huang-Rhys factor $S=2$
(right panels). The results are shown for three different scaled
temperatures $T_{\omega}$.} \label{fig:Morse_anhar}
\end{figure}

\section{$\bold{\text{SeO}_{2}+e^{-}\leftarrow\text{SeO}_{2}^{-}}$ photoelectron spectrum~\label{Sec:SeO2}}

Equilibrium nuclear geometries in the ground electronic states of $\text{SeO}_{2}^{-}$ (point group $C_{2v}$) and $\text{SeO}_{2}$ (point group $C_{2v}$) were calculated using the BPW91 functional and the aug-cc-pVQZ basis set. The computed values are provided in Table~\ref{tab:geom_SeO2} ($r$ is the SeO bond length, $\theta$ is the OSeO angle). 
Table~\ref{tab:freq_SeO2} presents the harmonic vibrational wavenumbers and the dimensionless Huang-Rhys displacement parameter $S$ along the normal-mode coordinates. 
All our calculations automatically account for Duschinsky effect, which is reflected in the non-diagonal elements of the excited-state Hessian matrix represented in the ground-state normal modes, because a single set of coordinates is used to represent all quantities (see Sec. 4 of the main text).
The shifts applied to the computed spectra are reported in Table~\ref{tab:shift_SeO2}.

\begin{table}[h]
 \caption{Equilibrium geometries of $\text{SeO}_{2}^{-}$ and $\text{SeO}_{2}$ at the BPW91/aug-cc-pVQZ level of theory.}
    \label{tab:geom_SeO2}
    \centering
    \begin{tabular}{cdd}
        \toprule
         &  \multicolumn{1}{c}{$r/\text{\r{A}}$} & \multicolumn{1}{c}{$\theta/\text{deg}$} \\
         \hline
         $\text{SeO}_{2}^{-}\,(\tilde{\text{X}}\,^{2}\text{B}_{1})$ & 1.6988 & 112.12 \\
         $\text{SeO}_{2}\,(\tilde{\text{X}}\,^{1}\text{A}_{1})$ & 1.6296 & 114.21 \\
         \botrule
    \end{tabular}
\end{table}

\begin{table}[h]
 \caption{Computed vibrational wavenumbers of $\text{SeO}_{2}^{-}$ ($\tilde{\nu}_{j}^{\prime\prime}$) and $\text{SeO}_{2}$ ($\tilde{\nu}_{j}^{\prime}$), and Huang-Rhys parameters $S_{j}$ between these two species at the BPW91/aug-cc-pVQZ level of theory.}
    \label{tab:freq_SeO2}
    \centering
    \begin{tabular}{ccddd}
        \toprule
        Mode label $j$ & Symmetry & \multicolumn{1}{c}{$\tilde{\nu}_{j}^{\prime\prime}/\text{cm}^{-1}$} & \multicolumn{1}{c}{$\tilde{\nu}_{j}^{\prime}/\text{cm}^{-1}$} & \multicolumn{1}{c}{$S_{j}$}\\
        \hline 
        1 & a${_1}$ & 778.57 & 914.54 & 1.62 \\
        2 & a${_1}$ & 311.25 & 350.79 & 0.03 \\
        3 & b${_2}$ & 788.35 & 951.94 & 0 \\
        \botrule
    \end{tabular}
\end{table}

\begin{figure} 
\includegraphics{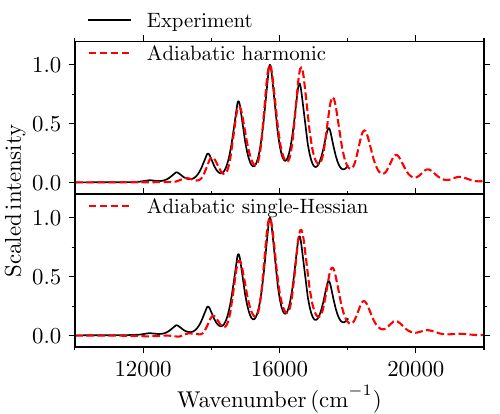}
\caption{Photoelectron spectrum of $\text{SeO}_{2}^{-}$ at $700\,\text{K}$. Spectra computed from the adiabatic harmonic and adiabatic single-Hessian coherence thermofield Gaussian wavepacket dynamics are compared to the experimental spectrum.~\cite{Snodgrass_Bowen:1989}}
\label{fig:SeO2_anharmonicity}%
\end{figure}

In Fig.~3 of the main text, we showed that single-Hessian GWD outperforms the global harmonic model in reproducing the experimental spectrum of $\text{SeO}_{2}^{-}$ when both methods are constructed using the vertical Hessian. Here, we perform an analogous comparison using the adiabatic Hessian.
Figure~\ref{fig:SeO2_anharmonicity} shows that adiabatic single-Hessian GWD captures the intensity of the high-frequency peaks better than the adiabatic harmonic model.
Comparison of Fig.~3 of the main text and Fig.~\ref{fig:SeO2_anharmonicity} highlights that the choice of the reference geometry has only a minor impact on single-Hessian results, whereas it greatly affects the accuracy of the global harmonic approximation.

\begin{table}
    \caption{Energy shifts applied to the computed photoelectron spectra of $\text{SeO}_{2}^{-}$.}
    \label{tab:shift_SeO2}
    \centering
    \begin{tabular}{ld}
        \toprule
         Variant of GWD & \multicolumn{1}{c}{Energy shift / $\text{cm}^{-1}$} \\
         \hline
         Adiabatic harmonic & -83 \\ 
         Vertical harmonic & -185\\ 
         Adiabatic single-Hessian & -83\\ 
         Vertical single-Hessian & 122 \\ 
         Local harmonic & -49 \\ 
         \botrule
    \end{tabular}
\end{table}

Single-Hessian GWD was performed using a fourth-order geometric integrator. Here, we measure the accuracy and efficiency of such integrator against the basic second-order integrator for convergence of the vibronic spectrum.
Fourth-order geometric integrators can be constructed using different composition schemes.~\cite{Choi_Vanicek:2019} For example, the triple-jump composition requires three steps of the second-order integrator, whereas Suzuki's fractal composition requires five. In this work, we used the latter, which is regarded as the optimal fourth-order composition because it minimizes the magnitude of the composition coefficients.~\cite{Choi_Vanicek:2019} 
Therefore, we compare the fourth-order integrator with time step $80 \, \text{a.u.}$ to the second-order integrator with time step $16 \, \text{a.u.}$ In this way, the computational cost of the two integrators is equivalent.
For a time step $\Delta t$, the convergence error is evaluated as the $L^{2}$-distance $\lVert\sigma^{(\Delta t)}-\sigma^{(\Delta t/2)}\rVert/\lVert\sigma^{(\Delta t/2)}\rVert$, where $\sigma^{(\Delta t)}$ is the spectrum obtained from the autocorrelation function [Eq.~(4) of the main text] evaluated using a time step $\Delta t$ and $\sigma^{(\Delta t/2)}$ is the spectrum obtained from the autocorrelation function evaluated every two steps of size $\Delta t/2$.~\cite{Barbiero_Vanicek:2026}
For a time step of $16 \, \text{a.u.}$, the second-order integrator achieves an error of $9.3\times10^{-4}$, whereas for a time step of $80 \, \text{a.u.}$, the fourth-order integrator achieves an error of $7.1\times10^{-5}$. This confirms the improved accuracy and efficiency of the fourth-order integrator.~\cite{Barbiero_Vanicek:2026}


\section{Absorption spectrum of phenyl radical}

Equilibrium nuclear geometries in the ground [$\text{D}_{0}\,(\tilde{X}\,^{2}\text{A}_{1})$, point group $C_{2v}$] and first excited [$\text{D}_{1}\,(\tilde{A}\,^{2}\text{B}_{1})$, point group $C_{2v}$] electronic states of phenyl radical were calculated using the CAM-B3LYP functional and the 6-31+G** basis set. The optimized structures in Cartesian coordinates are provided in Tables~\ref{tab:ground_phe} and~\ref{tab:excited_phe}. Table~\ref{tab:freq_phe} presents the harmonic vibrational wavenumbers, while the shifts applied to the computed spectra are reported in Table~\ref{tab:shift_phe}. 
The adiabatic harmonic spectrum was shifted to match the 0-0 transition, because this peak is uniquely defined in both the experimental and computed spectra, while all other computed spectra were shifted to match the most intense experimental peak.

\begin{table}[h]
    \caption{Ground-state equilibrium geometry (in $\text{\r{A}}$) of phenyl at the\\ CAM-B3LYP/6-31+G** level of theory.}
    \label{tab:ground_phe}
    \centering
    \begin{tabular}{cddd}
         \toprule
         &  \multicolumn{1}{c}{$x$} & \multicolumn{1}{c}{$y$} & \multicolumn{1}{c}{$z$}\\
         \hline
C & 0.0000 & 0.0000 & 1.3937 \\
C & 0.0000 & 1.2240 & 0.7695 \\
C & 0.0000 & 1.2108 & -0.6300 \\
C & 0.0000 & 0.0000 & -1.3202 \\
C & 0.0000 & -1.2108 & -0.6300 \\
C & 0.0000 & -1.2240 & 0.7695 \\
H & 0.0000 & 2.1582 & 1.3217 \\
H & 0.0000 & 2.1494 & -1.1764 \\
H & 0.0000 & 0.0000 & -2.4054 \\
H & 0.0000 & -2.1494 & -1.1764 \\
H & 0.0000 & -2.1582 & 1.3217 \\
         \botrule
    \end{tabular}
\end{table}
\begin{table}[]
    \caption{Excited-state equilibrium geometry (in $\text{\r{A}}$) of phenyl at the\\ TD-CAM-B3LYP/6-31+G** level of theory.}
    \label{tab:excited_phe}
    \centering
    \begin{tabular}{cddd}
        \toprule
         &  \multicolumn{1}{c}{$x$} & \multicolumn{1}{c}{$y$} & \multicolumn{1}{c}{$z$}\\
         \hline
C & 0.0000 & 0.0000 & 1.5409 \\
C & 0.0000 & 1.2104 & 0.7328 \\
C & 0.0000 & 1.2227 & -0.6443 \\
C & 0.0000 & 0.0000 & -1.3327 \\
C & 0.0000 & -1.2227 & -0.6443 \\
C & 0.0000 & -1.2104 & 0.7328 \\
H & 0.0000 & 2.1640 & 1.2565 \\
H & 0.0000 & 2.1545 & -1.2025 \\
H & 0.0000 & 0.0000 & -2.4193 \\
H & 0.0000 & -2.1545 & -1.2025 \\
H & 0.0000 & -2.1640 & 1.2565 \\
         \botrule
    \end{tabular}
\end{table}

\begin{table}[]
    \caption{Computed vibrational wavenumbers in $\text{D}_{0}$ ($\tilde{\nu}_{j}^{\prime\prime}$) and $\text{D}_{1}$ ($\tilde{\nu}_{j}^{\prime}$) electronic states of phenyl radical, and Huang-Rhys parameters $S_{j}$ between the corresponding two surfaces at the CAM-B3LYP/6-31+G** level of theory.}
    \label{tab:freq_phe}
    \centering
    \begin{tabular}{ccddd}
        \toprule
         Mode label $j$ & Symmetry & \multicolumn{1}{c}{$\tilde{\nu}_{j}^{\prime\prime}/\text{cm}^{-1}$} & \multicolumn{1}{c}{$\tilde{\nu}_{j}^{\prime}/\text{cm}^{-1}$} & \multicolumn{1}{c}{$S_{j}$}\\
        \hline
        1 & a${_1}$ & 3230.85 & 3223.65 & 0.00 \\
        2 & a${_1}$ & 3220.02 & 3203.28 & 0.00 \\
        3 & a${_1}$ & 3201.09 & 3185.88 & 0.00 \\
        4 & a${_1}$ & 1619.91 & 1685.03 & 0.50 \\
        5 & a${_1}$ & 1497.19 & 1472.86 & 0.28 \\
        6 & a${_1}$ & 1186.50 & 1228.96 & 0.02 \\
        7 & a${_1}$ & 1066.11 & 1051.93 & 0.37 \\
        8 & a${_1}$ & 1036.27 & 1015.35 & 0.11 \\
        9 & a${_1}$ & 992.28 & 962.95 & 2.07\\
        10 & a${_1}$ & 622.78 & 601.15 & 1.02\\
        11 & a${_2}$ & 991.12 & 1012.15 & 0 \\
        12 & a${_2}$ & 835.07 & 806.09 & 0 \\
        13 & a${_2}$ & 407.63 & 303.68 & 0 \\
        14 & b${_1}$ & 1017.53 & 1047.12 & 0 \\
        15 & b${_1}$ & 912.72 & 1004.40 & 0 \\
        16 & b${_1}$ & 733.00 & 737.45 & 0 \\
        17 & b${_1}$ & 677.57 & 654.83 & 0 \\
        18 & b${_1}$ & 431.83 & 354.64 & 0 \\
        19 & b${_2}$ & 3222.94 & 3211.60 & 0 \\
        20 & b${_2}$ & 3207.52 & 3186.89 & 0 \\
        21 & b${_2}$ & 1674.13 & 1572.42 & 0 \\
        22 & b${_2}$ & 1483.75 & 1417.03 & 0 \\
        23 & b${_2}$ & 1338.64 & 1339.76 & 0 \\
        24 & b${_2}$ & 1319.85 & 1270.44 & 0 \\
        25 & b${_2}$ & 1183.42 & 1152.31 & 0 \\
        26 & b${_2}$ & 1088.93 & 1080.20 & 0 \\
        27 & b${_2}$ & 604.61 & 562.07 & 0 \\
        \botrule
    \end{tabular}
\end{table}

\begin{table}[]
    \caption{Energy shifts applied to the computed absorption spectra of phenyl radical.}
    \label{tab:shift_phe}
    \centering
    \begin{tabular}{ld}
        \toprule
         Variant of GWD & \multicolumn{1}{c}{Energy shift / $\text{cm}^{-1}$} \\
         \hline
         Adiabatic harmonic & -160 \\
         Vertical harmonic & -177 \\
         Adiabatic single-Hessian & -194 \\
         Vertical single-Hessian & -61 \\
         \botrule
    \end{tabular}
\end{table}

\begin{figure*} [h]
\includegraphics{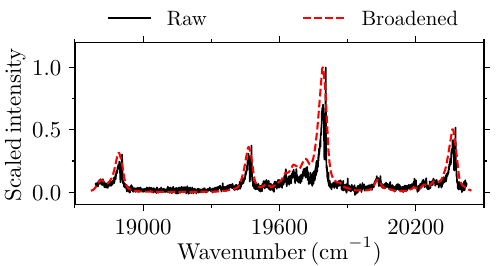}
\caption{Absorption spectrum of phenyl radical at $298\,\text{K}$. The broadened spectrum is compared to the raw experimental spectrum.~\cite{Tonokura_Wallington:2002}}
\label{fig:phe_exp}%
\end{figure*}

In Fig.~4 of the main text, the rotationally resolved experimental spectrum~\cite{Tonokura_Wallington:2002} of phenyl radical was convolved with a Lorentzian with half width at half maximum of $8.45\, \text{cm}^{-1}$ to remove rotational resolution. 
Before broadening, a function $f(\omega)=\text{min}[\sigma(\omega),5\,\text{ppm}]$ was subtracted from the spectrum $\sigma(\omega)$ to remove noise.
The raw and broadened spectra are shown in Fig.~\ref{fig:phe_exp}.   

\begin{figure*} 
\includegraphics{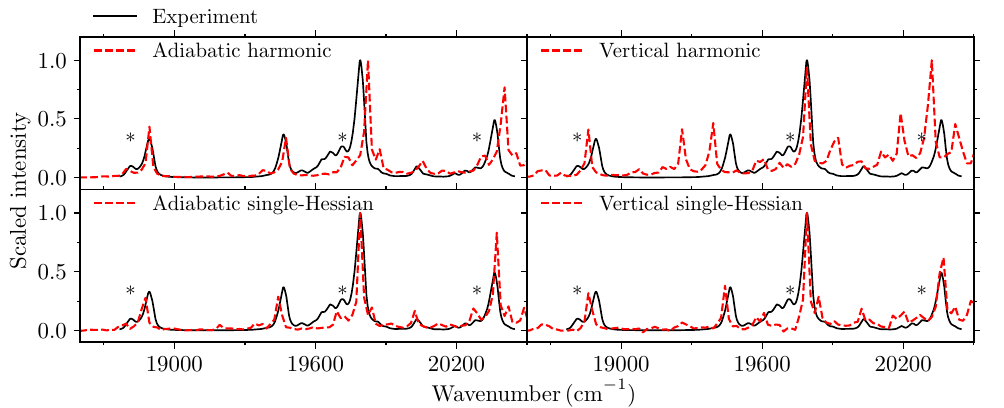}
\caption{Absorption spectrum of phenyl radical at $298\,\text{K}$. Spectra computed from the adiabatic and vertical harmonic, and from the adiabatic and vertical single-Hessian coherence thermofield Gaussian wavepacket dynamics are compared to the experimental spectrum.~\cite{Tonokura_Wallington:2002} Hot bands are denoted by asterisks.}
\label{fig:phe_anharmonicity}%
\end{figure*}

In Fig.~\ref{fig:phe_anharmonicity}, we compare the adiabatic and vertical single-Hessian spectra with the adiabatic and vertical harmonic ones.
In contrast to global harmonic models, the single-Hessian method recovers the correct spacing of the peaks, independently of the choice of the reference Hessian.
Interestingly, the vertical single-Hessian calculation yields more accurate peak intensities, whereas the adiabatic single-Hessian method captures the position of hot bands better.

\begin{figure*} 
\includegraphics{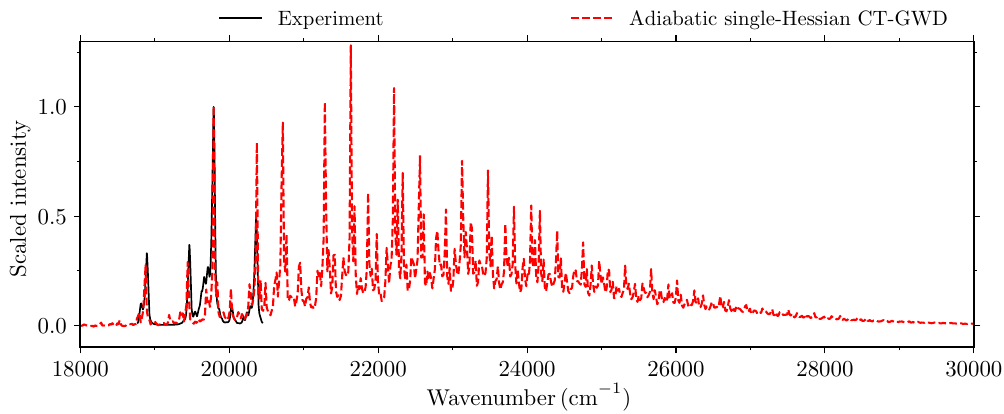}
\caption{Absorption spectrum of phenyl radical at $298\,\text{K}$. The spectrum computed from the adiabatic single-Hessian coherence thermofield Gaussian wavepacket dynamics is compared to the experimental spectrum.~\cite{Tonokura_Wallington:2002}}
\label{fig:Phe_full_spec}%
\end{figure*}

Finally, in Fig.~\ref{fig:Phe_full_spec} we report the whole frequency range of the rotationally averaged $\text{D}_{1} \leftarrow \text{D}_{0}$ spectrum computed from the adiabatic single-Hessian coherence thermofield GWD.

\section{Absorption spectrum of naphthalene}

Equilibrium nuclear geometries in the ground [$\text{S}_{0}\,(\tilde{\text{X}}\,\text{A}_{\text{g}})$, point group $D_{2h}$] and first excited [$\text{S}_{1}\,(\text{B}_{2\text{u}})$, point group $D_{2h}$] electronic states of naphthalene were calculated using the B3LYP functional and the 6-31+G** basis set. The computed first excited state corresponds to the second excited state ($\tilde{\text{B}}\,\text{B}_{2\text{u}}$) observed experimentally.~\cite{George_Morris:1968,Rubio_Roos:1994}
The optimized structures in Cartesian coordinates are provided in Tables~\ref{tab:ground_naph} and~\ref{tab:excited_naph}. The molecule has been placed in the $xy$ plane with the long molecular
axis corresponding to the $x$ axis.

\begin{table}[h]
    \caption{Ground-state equilibrium geometry (in $\text{\r{A}}$) of naphthalene at the B3LYP/6-31+G** level of theory.}
    \label{tab:ground_naph}
    \centering
    \begin{tabular}{cddd}
         \toprule
         &  \multicolumn{1}{c}{$x$} & \multicolumn{1}{c}{$y$} & \multicolumn{1}{c}{$z$}\\
         \hline
C & 2.4363 & 0.7091 & 0.0000 \\
C & 1.2459 & 1.4039 & 0.0000 \\
C & 0.0000 & 0.7173 & 0.0000 \\
C & -1.2459 & 1.4039 & 0.0000 \\
C & -2.4363 & 0.7091 & 0.0000 \\
C & -2.4363 & -0.7091 & 0.0000 \\
C & -1.2459 & -1.4039 & 0.0000 \\
C & 0.0000 & -0.7173 & 0.0000 \\
C & 1.2459 & -1.4039 & 0.0000 \\
C & 2.4363 & -0.7091 & 0.0000 \\
H & 3.3803 & 1.2464 & 0.0000 \\
H & 1.2443 & 2.4910 & 0.0000 \\
H & -1.2443 & 2.4910 & 0.0000 \\
H & -3.3803 & 1.2464 & 0.0000 \\
H & -3.3803 & -1.2464 & 0.0000 \\
H & -1.2443 & -2.4910 & 0.0000 \\
H & 1.2443 & -2.4910 & 0.0000 \\
H & 3.3803 & -1.2464 & 0.0000 \\
         \botrule
    \end{tabular}
\end{table}
\begin{table}[]
    \caption{Excited-state equilibrium geometry (in $\text{\r{A}}$) of naphthalene at the\\ TD-B3LYP/6-31+G** level of theory.}
    \label{tab:excited_naph}
    \centering
    \begin{tabular}{cddd}
        \toprule
         &  \multicolumn{1}{c}{$x$} & \multicolumn{1}{c}{$y$} & \multicolumn{1}{c}{$z$}\\
         \hline
C  &  2.4864  &  0.6901  &  0.0000  \\
C  &  1.2436  &  1.3992  &  0.0000  \\
C  &  0.0000  &  0.7206  &  0.0000  \\
C  &  -1.2436  &  1.3992  &  0.0000  \\
C  &  -2.4864  &  0.6901  &  0.0000  \\
C  &  -2.4864  &  -0.6901  &  0.0000  \\
C  &  -1.2436  &  -1.3992  &  0.0000  \\
C  &  0.0000  &  -0.7206  &  0.0000  \\
C  &  1.2436  &  -1.3992  &  0.0000  \\
C  &  2.4864  &  -0.6901  &  0.0000  \\
H  &  3.4181  &  1.2468  &  0.0000  \\
H  &  1.2498  &  2.4858  &  0.0000  \\
H  &  -1.2498  &  2.4858  &  0.0000  \\
H  &  -3.4181  &  1.2468  &  0.0000  \\
H  &  -3.4181  &  -1.2468  &  0.0000  \\
H  &  -1.2498  &  -2.4858  &  0.0000  \\
H  &  1.2498  &  -2.4858  &  0.0000  \\
H  &  3.4181  &  -1.2468  &  0.0000  \\
         \botrule
    \end{tabular}
\end{table}

\begin{table}[]
    \caption{Energy shifts ($\text{cm}^{-1}$) applied to the computed $\text{S}_{1}\leftarrow\text{S}_{0}$ spectra of naphthalene.}
    \label{tab:shift_naph}
    \centering
    \begin{tabular}{lcccc}
        \toprule
         Variant of GWD & 297 K & 423 K & 573 K & 773 K \\
         \hline
         Adiabatic single-Hessian coherence thermofield GWD & \multicolumn{4}{c}{5156}\\
         Adiabatic single-Hessian GWD 0 K & 5090 & 5023 & 4907 & 4791\\
         \botrule
    \end{tabular}
\end{table}

\begin{table}[]
    \caption{Half width at half maximum ($\text{cm}^{-1}$) of Gaussian broadening function applied to the computed absorption spectra of naphthalene.}
    \label{tab:broad_naph}
    \centering
    \begin{tabular}{lcccc}
        \toprule
         Method & 297 K & 423 K & 573 K & 773 K \\
         \hline
         Adiabatic single-Hessian coherence thermofield GWD & \multicolumn{4}{c}{50} \\
         Adiabatic single-Hessian GWD (0 K) & 109 & 190 & 338 & 608\\
         \botrule
    \end{tabular}
\end{table}

\begin{figure*} 
\includegraphics{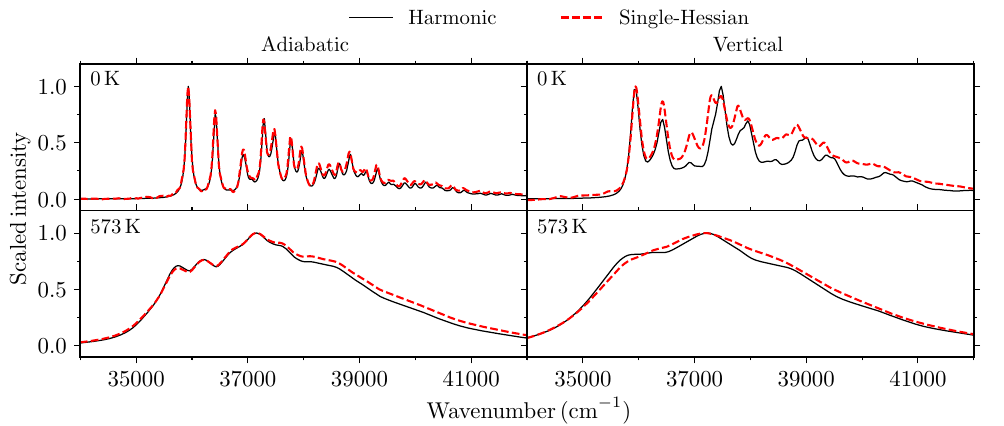}
\caption{Absorption spectra of naphthalene at two different temperatures. Spectra computed from the single-Hessian coherence thermofield Gaussian wavepacket dynamics are compared to the harmonic spectra for two choices of the reference Hessian (adiabatic and vertical).}
\label{fig:naph_sha_vs_gha}%
\end{figure*}

Due to its rigid structure, naphthalene is expected to display minimal anharmonicity. This is supported by Fig.~\ref{fig:naph_sha_vs_gha}, where the spectra obtained from the adiabatic single-Hessian coherence thermofield GWD closely match those from the adiabatic harmonic model. However, the presence of an imaginary frequency in the Hessian leads to nonphysical broadening in both the vertical harmonic and vertical single-Hessian GWD spectra.

\begin{figure*} 
\includegraphics{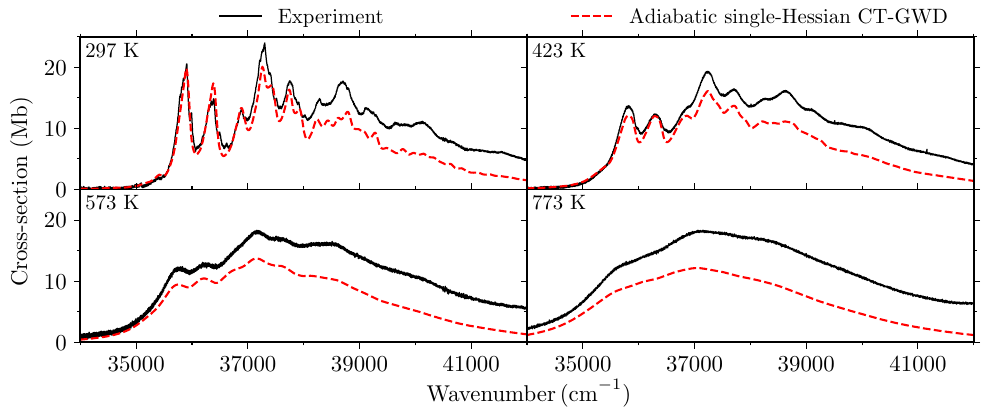}
\caption{Absolute absorption cross-sections of naphthalene at different temperatures. Cross-sections computed from the adiabatic single-Hessian coherence thermofield Gaussian wavepacket dynamics (CT-GWD) are compared to the experimental cross-sections.~\cite{Grosch_Fateev:2015}}
\label{fig:naph_abs_spectra}%
\end{figure*}

Figure~\ref{fig:naph_abs_spectra} compares the simulated and experimental absolute cross-sections at four different temperatures. The spectra computed from the adiabatic single-Hessian coherence thermofield GWD slightly underestimate the experimental values. 
Several factors probably contribute to the observed deviation: (i)~inaccuracies in the electronic structure calculations of the transition dipole moment, (ii)~neglect of Herzberg–Teller effect, and (iii)~neglect of transitions to higher-excited electronic states whose absorption bands overlap with that of the target state at higher temperatures.

\section{Absorption spectrum of aminocoumarin C450}

Equilibrium nuclear geometries in the ground ($\text{S}_{0}$, point group $C_{1}$) and first excited ($\text{S}_{1}$, point group $C_{s}$) electronic states of aminocoumarin C450 were calculated using the $\omega$B97X-D functional and the 6-31+G** basis set. The optimized structures in Cartesian coordinates are provided in Tables~\ref{tab:ground_C450} and~\ref{tab:excited_C450}. The shifts applied the computed spectra are reported in Table~\ref{tab:shift_C450}.

\begin{table}[h]
    \caption{Ground-state equilibrium geometry (in $\text{\r{A}}$) of aminocoumarin C450 at the $\omega$B97X-D/6-31+G** level of theory.}
    \label{tab:ground_C450}
    \centering
    \begin{tabular}{cddd|cddd}
         \toprule
         & \multicolumn{1}{c}{$x$} & \multicolumn{1}{c}{$y$} & \multicolumn{1}{c}{$z$} & & \multicolumn{1}{c}{$x$} & \multicolumn{1}{c}{$y$} & \multicolumn{1}{c}{$z$}\\
         \hline
O & 1.4470 & -1.6000 & -0.0233 &
O & 3.4972 & -2.4413 & -0.0035 \\
N & -3.0555 & -0.0900 & -0.0928 &
C & 1.0572 & 0.7772 & 0.0017 \\
C & -1.7076 & 0.1821 & -0.0405 &
C & -1.2694 & 1.5383 & -0.0193 \\
C & 0.5932 & -0.5413 & -0.0217 &
C & 0.0859 & 1.7962 & 0.0033 \\
C & -0.7628 & -0.8473 & -0.0403 &
C & 2.4888 & 0.9992 & 0.0236 \\
C & -2.2761 & 2.6579 & -0.0295 &
C & -3.5930 & -1.4279 & 0.0643 \\
C & 3.3128 & -0.0753 & 0.0210 &
C & 3.0314 & 2.3991 & 0.0475 \\
C & 2.8176 & -1.4421 & -0.0018 &
C & -5.1141 & -1.3935 & 0.0440 \\
H & 0.4130 & 2.8321 & 0.0193 &
H & -1.0493 & -1.8913 & -0.0521 \\
H & -3.6730 & 0.6596 & 0.1760 &
H & -2.9307 & 2.6017 & -0.9076 \\
H & -2.9148 & 2.6440 & 0.8637 &
H & -1.7757 & 3.6285 & -0.0509 \\
H & -3.2301 & -2.0551 & -0.7587 &
H & -3.2386 & -1.8849 & 1.0010 \\
H & 4.3916 & 0.0251 & 0.0363 &
H & 2.6770 & 2.9380 & 0.9325 \\
H & 2.7001 & 2.9590 & -0.8333 &
H & 4.1228 & 2.3948 & 0.0618 \\
H & -5.5075 & -0.7973 & 0.8747 &
H & -5.4819 & -0.9678 & -0.8944 \\
H & -5.5164 & -2.4046 & 0.1418 \\
         \botrule
    \end{tabular}
\end{table}
\begin{table}[h]
    \caption{Excited-state equilibrium geometry (in $\text{\r{A}}$) of aminocoumarin C450 at the TD-$\omega$B97X-D/6-31+G** level of theory.}
    \label{tab:excited_C450}
    \centering
    \begin{tabular}{cddd|cddd}
         \toprule
         & \multicolumn{1}{c}{$x$} & \multicolumn{1}{c}{$y$} & \multicolumn{1}{c}{$z$} & & \multicolumn{1}{c}{$x$} & \multicolumn{1}{c}{$y$} & \multicolumn{1}{c}{$z$}\\
         \hline
O  &  0.2148  &  -2.2529  &  0.0000 & 
O  &  -0.5369  &  -4.3488  &  0.0000  \\
N  &  2.0861  &  2.0869  &  0.0000 & 
C  &  -1.3394  &  -0.3683  &  0.0000  \\
C  &  0.9843  &  1.2845  &  0.0000  & 
C  &  -0.3238  &  1.8782  &  0.0000 \\
C  &  0.0000  &  -0.9270  &  0.0000  & 
C  &  -1.4333  &  1.0538  &  0.0000  \\
C  &  1.1158  &  -0.1202  &  0.0000  & 
C  &  -2.4523  &  -1.2445  &  0.0000  \\
C  &  -0.4743  &  3.3730  &  0.0000  & 
C  &  3.4546  &  1.6106  &  0.0000  \\
C  &  -2.1833  &  -2.6299  &  0.0000  & 
C  &  -3.8506  &  -0.7142  &  0.0000  \\
C  &  -0.8751  &  -3.1787  &  0.0000  & 
C  &  4.4265  &  2.7811  &  0.0000  \\
H  &  -2.4194  &  1.5059  &  0.0000  & 
H  &  2.0864  &  -0.5999  &  0.0000  \\
H  &  1.9480  &  3.0849  &  0.0000  & 
H  &  -0.0148  &  3.8278  &  0.8871  \\
H  &  -0.0148  &  3.8278  &  -0.8871  & 
H  &  -1.5292  &  3.6535  &  0.0000  \\
H  &  3.6244  &  0.9805  &  0.8838  & 
H  &  3.6244  &  0.9805  &  -0.8838  \\
H  &  -2.9898  &  -3.3541  &  0.0000 & 
H  &  -4.0505  &  -0.0912  &  -0.8833 \\
H  &  -4.0505  &  -0.0912  &  0.8833  & 
H  &  -4.5744  &  -1.5325  &  0.0000  \\
H  &  4.2928  &  3.4047  &  -0.8898  & 
H  &  4.2928  &  3.4047  &  0.8898  \\
H  &  5.4548  &  2.4131  &  0.0000  \\
         \botrule
    \end{tabular}
\end{table}

\begin{table}[h]
    \caption{Energy shifts applied to the computed absorption spectra of aminocoumarin C450.}
    \label{tab:shift_C450}
    \centering
    \begin{tabular}{ld}
        \toprule
         Variant of GWD & \multicolumn{1}{c}{Energy shift / $\text{cm}^{-1}$} \\
         \hline
         Adiabatic harmonic & -2120 \\
         Vertical harmonic & -1956\\
         Adiabatic single-Hessian & -1792\\
         Vertical single-Hessian & -1727 \\
         \botrule
    \end{tabular}
\end{table}

\begin{figure}
\includegraphics{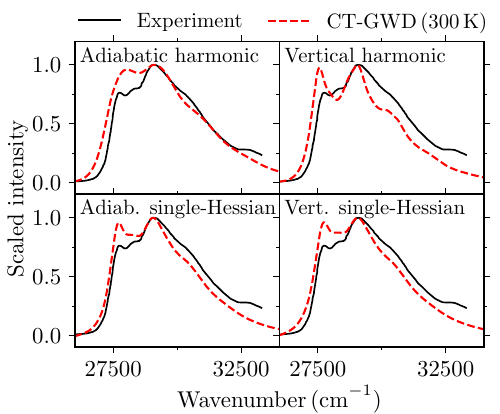} \caption{Absorption
spectrum of aminocoumarin C450 at $300\,\text{K}$. Spectra computed from
either the adiabatic and vertical harmonic, or the adiabatic and vertical
single-Hessian coherence thermofield Gaussian wavepacket dynamics (CT-GWD)
are compared to the experimental spectrum.~\cite{MunizMiranda_Barone:2015}} %
\label{fig:C450_sha_vs_gha}
\end{figure}

In Fig.~3 of the main text, we highlighted the importance of accounting for anharmonicity in simulation of medium-resolution spectra. Fig.~\ref{fig:C450_sha_vs_gha} highlights the impact of
this effect even in low-resolution spectra.
For different reasons,~\cite{Begusic_Vanicek:2019,Barbiero_Vanicek:2026}
both vertical and adiabatic harmonic calculations fail to reproduce the
shape of the experimental spectrum. In contrast, such shape is captured by
the single-Hessian method, independently of the choice (adiabatic or
vertical) of the reference Hessian.

\bibliographystyle{aipnum4-2}
\bibliography{Single_Hessian_thermofield_v34}